\documentclass[amssymb,aps,twocolumn,floats]{revtex4}
\usepackage{psfig}
\usepackage{color,graphicx,pstricks}
\draft
\begin{document}
\title{Focussing Quantum States}
\author{M. Sentef, A. P. Kampf, S. Hembacher, and J. Mannhart}
\affiliation{Institute of Physics, Center for Electronic Correlations and Magnetism,\\
University of Augsburg, 86135 Augsburg, Germany}

\begin{abstract}
Does the size of atoms present a lower limit to the size of electronic 
structures that can be fabricated in solids? This limit can be overcome 
by using devices that exploit quantum mechanical scattering of electron waves 
at atoms arranged in focussing geometries on selected surfaces. Calculations 
reveal that features smaller than a hydrogen atom can be obtained. These 
structures are potentially useful for  device applications and offer a route 
to the fabrication of ultrafine and well defined tips for scanning tunneling 
microscopy.
\vskip0.2cm
\noindent PACS numbers: 73.20.-r, 73.63.-b, 85.35.-p
\end{abstract}
\vskip0.15cm

\maketitle

The manufacture of ever smaller objects is an ongoing pursuit of science and 
technology, which at the end of the 20th century led to the fabrication of 
nanometer-sized structures. A seminal highlight was accomplished in 1993 with 
the manipulation of single atoms \cite{Eigler}, which were even assembled into 
crystallites \cite{molecules}. It obviously seems prohibited to construct 
even smaller structures. How could this be done?

Here, we explore the possibility to design ultrasmall electronic structures by 
manipulating electronic surface states of metals. We will present examples 
revealing that electron density peaks as small as 1 {\AA} can be achieved. The 
width of the electronic peak is hereby limited only on the scale of the 
shortest wavelength of the surface band states. By shrinking the size of 
interference peaks of electronic surface states, new options for device 
application arise. Electron density peaks of {\AA}-width may for example be 
exploited as ultrafine and well defined quantum states, to be used as tips in 
scanning tunneling microscopy (STM).

The approach discussed below builds on experimental investigations of 
electronic surface states. Electrons in Shockley surface states of metals can 
be scattered by surface steps and by individual atoms placed on the surface 
\cite{Eigler,Hasegawa,Buergi}. Complex interference patterns have been 
generated in artificially manufactured corrals of circular or elliptical shape
\cite{Heller,Kliewer}. Even quantum mirage phenomena have been induced in 
such corrals \cite{Manoharan,Fiete,Aligia}. In quantum corrals, electrons are 
focussed on well defined areas on the surface, thereby creating locations with 
an enhanced local density of states and therefore an enhanced electron density 
with typical sizes of 1-2 nm. This work has opened a route for manipulating 
quantum states almost on the atomic level and raises the question whether it 
is possible to design arrangements of atoms with optimized focussing 
properties for quantum waves. Can quantum structures on sub-{\AA} lengthscales 
be realized? 

Fundamental as well as practical problems are encountered on the road to 
sharply focussed quantum states. First, one may ask whether Heisenberg's 
uncertainty principle \cite{Heisenberg} ultimately sets a limit for the 
spatial extent of fine structure in a quantum mechanical wavefunction. On the 
practical side, the rules of optics cannot be applied to design the focussing 
structures for quantum waves. This is because electronic waves with short 
wavelengths are needed to finely focus the electrons, but scattering of such 
high-energy particles involves anisotropic non--s-wave channels. Since the 
higher angular momentum 
scattering channels have no counterpart in classical wave mechanics, the 
design rules of conventional optical instruments cannot be used to device 
instruments for focussing quantum mechanical waves with short wavelengths.

\begin{figure}[t!]
\centerline{ \psfig{file=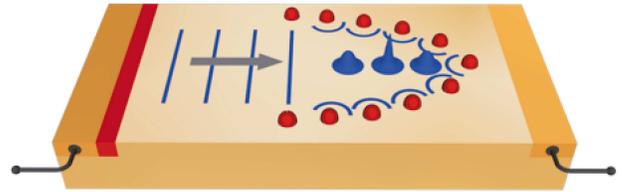,width=84mm,silent=}}
\caption{(Color online) Schematic view of the focussing geometry. An 
electronic surface wave is generated with a tunnel junction and propagates 
towards an arrangement of scattering centers (red semi-spheres). Multiple 
interference peaks emerge from the superposition of scattered waves.}
\label{sentef1}
\end{figure}

Using model calculations of surface wave scattering from hard spheres, we 
consider here focussing arrangements built from scattering centers (see Fig. 
\ref{sentef1}), designed to achieve ultra-narrow peak widths. Complex 
interference patterns are obtained and analyzed for parabolic and 
semi-elliptic geometries. It is shown that in this way locally enhanced 
electron densities with sub-{\AA} lateral size can be realized.

The guiding idea for our approach is to design quantum mechanical (electronic)
states $\Psi({\bf r},{\bf p})$ with effective widths $\Delta r$ and $\Delta p$
in real space and in momentum space, respectively, such that $|\Psi|^2$ forms 
a spike of width $\Delta r^*$. Heisenberg's uncertainty relation requires that
$\Delta r \Delta p \geq\hbar/2$, where $\hbar$ is Planck's constant. While 
this fundamental principle of quantum mechanics inevitably controls any 
measurement process, it is important that the uncertainty relation does not 
preclude the possibility to structure the electronic wavefunction on a 
lengthscale $\Delta r^*$ much smaller than $\Delta r$. Therefore, the 
principles of quantum mechanics do not set a lower limit for generating 
ultrasmall electronic structures, although these will possibly have a small 
local probability density in the spike volume $\Delta r^*$. Rather, in a 
superposition of quantum mechanical waves, $\Delta r^*$ is often limited by 
the largest available momentum, which thereby imposes an upper limit on 
$\Delta p$. For the purpose of focussing electronic waves in a crystalline 
solid this suggests to use high-energy waves preferentially in band states 
with a large effective mass.

To explore the size of the smallest area into which the electrons can be 
focussed with practical experimental setups, we performed model 
calculations in two space dimensions. Scattering centers of radius $r_0$ are 
arranged in open focussing geometries with either parabolic or semi-elliptic 
shape (see Fig. 1). An electronic surface wave, generated for example by a 
tunnel junction, is considered to enter the focussing arrangement as a plane 
wave with wavevector ${\bf k}$. The wave propagates along the symmetry axis of
a regular arrangement of hard disks, with which we model individual atoms 
placed on a metallic surface with a spacing $d\sim 10\, r_0$ as is typical 
for Fe adatom corrals \cite{Eigler,Heller}. For long wavelengths 
$\lambda\gg r_0$, realized for surface state electrons on copper (111) 
surfaces, only isotropic s-wave scattering is significant. In this case, 
multiple scattering events and absorption from the scattering centers can be 
straightforwardly considered \cite{Heller}. For shorter wavelengths,  
the established scattering analysis has to be extended to include 
higher angular momentum scattering channels.

In the absence of multiple scattering the scattering state has the
asymptotic form (for $kr\gg 1$) 
\begin{equation}
\psi({\bf r}) \simeq e^{{\rm i}{\bf k}\cdot{\bf r}}+\sum_{\nu}f(
\vartheta_{\nu})e^{{\rm i}{\bf k}\cdot{\bf R}_{\nu}}\frac{e^{{\rm i}kr_{\nu}}}
{\sqrt{kr_{\nu}}}\, ,
\label{asymp1}
\end{equation}
where $\textbf{R}_{\nu}$ denotes the position of the $\nu$-th scattering 
center, and $\textbf{r}_{\nu}={\bf r}-{\bf R}_\nu$ measures with the polar 
coordinates $r_{\nu}$ and $\vartheta_{\nu}$ the relative position to the disk 
at ${\bf R}_\nu$. Introducing partial wave phase shifts, the scattering 
amplitude follows as
\begin{equation}
f(\vartheta) = \sqrt{\frac{2{\rm i}}{\pi}} \left[e^{{\rm i}\delta_{0}}
\sin{\delta_{0}} + \sum_{m=1}^{\infty} 2 e^{{\rm i}\delta_{m}}
\sin{\delta_{m}} \cos{m\vartheta} \right].
\label{famplitude}
\end{equation}
The parameter $m$ counts the scattering channel; the corresponding phase 
shifts are determined by $\tan \delta_m=J_m(kr_0)/N_m(kr_0)$, where $J_{m}$ 
and $N_{m}$ denote the Bessel functions of the first and second kind, 
respectively.

In the restriction to s-wave scattering, repeated scattering events are 
included by extending Eq. (\ref{asymp1}) to
\begin{eqnarray}
\psi(\bf{r}) &\simeq& e^{{\rm i}{\bf k}\cdot{\bf r}} + {\bf b}_T\cdot
\left[{\bf 1}+{\bf A}+{\bf A}^2+{\bf A}^3+\ldots\right]\cdot {\bf a}({\bf r})
\nonumber \\
&=&e^{{\rm i}{\bf k}\cdot{\bf r}} + {\bf b}_T\cdot \left[{\bf 1}-{\bf A}
\right]^{-1}\cdot{\bf a}({\bf r}).
\end{eqnarray}
Here, ${\bf b}=(b_1,\ldots,b_N)$ for $N$ scattering centers with $b_\nu=e^{{
\rm i}{\bf k}\cdot{\bf R}_\nu}$ accounts for the phase factors related to the 
individual disk positions. The amplitudes for the waves scattered from the 
disk at ${\bf R}_\nu$ to the disk at ${\bf R}_\mu$ ($\nu\neq\mu$) form an 
$N\times N$ matrix with 
\begin{equation}
A_{\nu\mu}=f_{0} \; \frac{e^{ikr_{\nu\mu}}}{\sqrt{kr_{\nu\mu}}}\, ,
\end{equation}
where $r_{\nu\mu}=|{\bf R}_\nu-{\bf R}_\mu|$. Similarly, the amplitude of the 
wave scattered from ${\bf R}_\mu$ to ${\bf r}$ is 
\begin{equation}
a_{\mu}({\bf r})=f_0 \; \frac{e^{ikr_{\mu}}}{\sqrt{kr_{\mu}}}\, .
\end{equation}
The scattering amplitude $f_0$ is related to the s-wave phase shift $\delta_0$ 
by
\begin{eqnarray}
f_0 = \sqrt{\frac{2{\rm i}}{\pi}} \; e^{{\rm i}\delta_0}\sin\delta_0 = {1\over
\sqrt{2\pi{\rm i}}}\left(e^{2{\rm i}\delta_0}-1\right) \; .
\label{f0abs}
\end{eqnarray}
The possible partial absorption of the incident electronic surface wave by 
inelastic scattering and scattering into bulk states is incorporated by 
allowing the phase shifts to become complex \cite{Heller}, corresponding to
the replacement $e^{2{\rm i}\delta_0}\rightarrow\alpha_0 e^{2{\rm i}\delta_0}$
in Eq. (\ref{f0abs}). Henceforth, $\delta_0$ is a real number; the absorption 
coefficient $\alpha_0$ is 1 for non-absorbing adatoms and vanishes for 
complete attenuation.

For wavelengths which become almost comparable to the size of an atom, higher 
angular momentum scattering channels are important. To give an example, for 
$kr_0=2\pi r_0/\lambda=1.24$ (see below) the scattering phase shifts in s-, 
p-, and d-channels are $\delta_0=69^\circ$, $\delta_1=-41^\circ$, and $\delta_2
=-8^\circ$. With the restriction to double scattering from each disk the 
ansatz for the asymptotic scattering state is extended to 
\begin{eqnarray}
&&\psi({\bf r}) \simeq e^{{\rm i}{\bf k}\cdot{\bf r}} + \sum_{\nu=1}^N
e^{{\rm i}{\bf k}\cdot{\bf R}_\nu} \; f(\vartheta_\nu) \;
\frac{e^{{\rm i}kr_{\nu}}}{\sqrt{kr_{\nu}}}  \\ 
&+&\sum_{\mu ,\nu=1; \mu\neq \nu}^N e^{{\rm i}{\bf k}\cdot{\bf R}_{
\nu}} \; f(\vartheta_{\nu\mu}) \; \frac{e^{{\rm i}kr_{\nu\mu}}}{\sqrt{kr_{\nu
\mu}}} \;f(\vartheta_{\mu}-\vartheta_{\nu\mu}) \;
\frac{e^{{\rm i}kr_{\mu}}}{\sqrt{kr_{\mu}}}\, ,\nonumber
\label{psibe}
\end{eqnarray}
where $\vartheta_{\nu\mu}$ is the angle for the position of the scattering 
disk $\mu$ in the polar coordinate system attached to disk $\nu$. Without 
absorption, and neglecting the still small contribution of the d-wave 
scattering channel only, the s- and p-wave contributions ($m=0$ and $m=1$) 
are included in the angular dependent scattering amplitude given in Eq. 
(\ref{famplitude}). 

In a first attempt, the focussing properties of a device consisting of two 
parabolic ``quantum mirrors'' arranged like a reflector telescope, have been 
calculated. The substrate was assumed to be the Cu (111) surface, and 29 
hard disks with radius $r_0=0.63$ {\AA} were chosen to present Co$^{3+}$ ions
as scatterers. The focal distance of the parabola is $f=4.9$ {\AA}, and the 
average disk spacing is 8 {\AA}. The wavelength of the incoming wave was taken 
to be $\lambda=12$ {\AA}. At this wavelength $\lambda\gg r_0$, so that only 
s-wave scattering has to be considered. In Fig. 2 we show the resulting 
absolute square $|\psi({\bf r})|^2$ of the scattering state. Guided by the 
successful quantitative analysis of the current-voltage characteristics at 
the center of a circular quantum corral of iron atoms on a copper surface 
\cite{Heller}, the ``black dot'' attenuation limit $\alpha_0=0$ was adopted. 
The image shown in Fig. 2 is the pattern that would be observed in a 
standard STM local density of states measurement.

\begin{figure}[t!]
\centerline{ \psfig{file=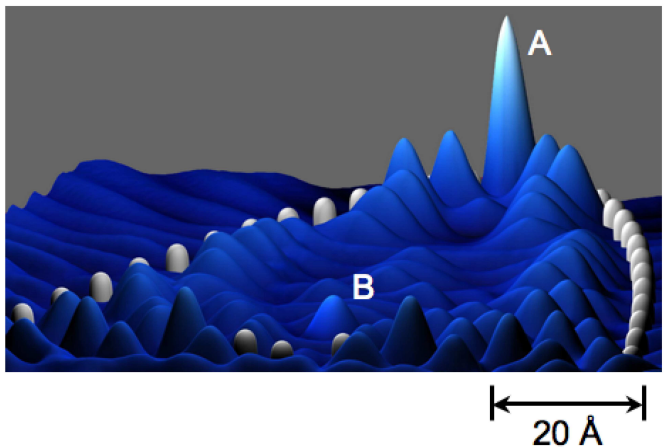,width=90mm,silent=}}
\caption{(Color online) Focussing of an electron wave with wavelength 12 {\AA}
by scattering from two quantum mirrors. The arrangement consists of a large 
parabolic mirror formed by 29 scattering centers (white pillars) and a small 
mirror consisting of 3 additional scatterers in a reflector telescope 
geometry. The plot shows the distribution $|\psi({\bf r})|^2$ of the 
electronic scattering state. Only s-wave scattering is included. ``A'' 
marks the most prominent peak near the tip, while the peak ``B'' emerges 
near the focus point as a result of the quantum mirror geometry.} 
\label{sentef2}
\end{figure}

Near the tip of the parabola intense interference peaks with a full width at 
half maximum (FWHM) $\sim 4.2$ {\AA} are produced (see for example peak A in 
Fig. \ref{sentef2}). Due to the $1/\sqrt{r}$ decay of the amplitude for the 
scattered waves the peak heights are larger the closer the peaks are to the 
scattering atom \cite{comment}. Resulting from the focussing of the second, 
smaller ``quantum mirror'' additional peaks emerge near its focal point (see 
for example peak B in Fig. \ref{sentef2}). The width of peak B, $\sim 3.5$ 
{\AA} at FWHM, is just fractions of the incoming wavelength. The peak, 
however, has a small intensity. 

There are obvious routes to further improve the focussing. First, materials 
capable of sustaining surface waves with considerably smaller wavelengths may 
be used. The goal to achieve interference peaks with subatomic widths 
precludes the use of surface eigenstates of noble metal surfaces, whose 
typical wavelengths are $\sim 15$ {\AA} \cite{Davison}. The recently observed 
Friedel oscillations on beryllium (0001) surfaces with wavelengths as short 
as 3.2 \AA \cite{Be} suggest Be as a candidate material. Other options for 
tuning the electronic density distribution include using non-monochromatic 
waves and optimizing the arrangement of the surface adatom scatterers and the 
geometry of the quantum mirror. The development of a mathematical algorithm 
to select a focussing arrangement is quite a non-trivial task, and we have 
therefore explicitly tried several device geometries. Of the ones explored, 
particularly sharp peaks were obtained by using a semi-ellipse, as we will 
demonstrate in the following.

\begin{figure}[t!]
\centerline{\psfig{file=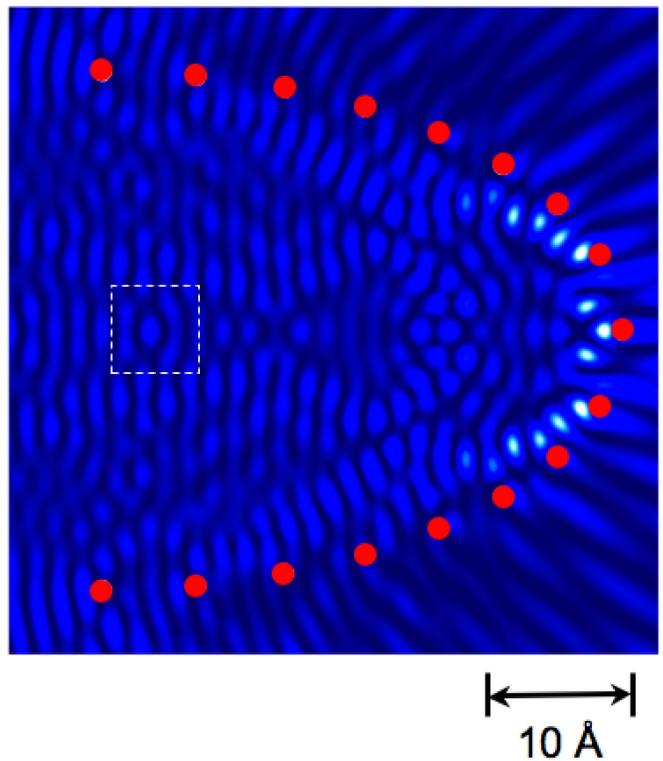,width=90mm,silent=}}
\caption{(Color online) Distribution of an electron scattering state 
$|\psi({\bf r})|^2$ achieved by scattering from a semi-elliptic arrangement. 
The wavelength of the incoming wave is $\lambda=3.2$ {\AA} -- the wavelength 
of Friedel oscillations on Be (0001) surfaces. s- and p-wave scattering 
channels are included.}
\label{sentef3}
\end{figure}

\begin{figure}[t!]
\centerline{\psfig{file=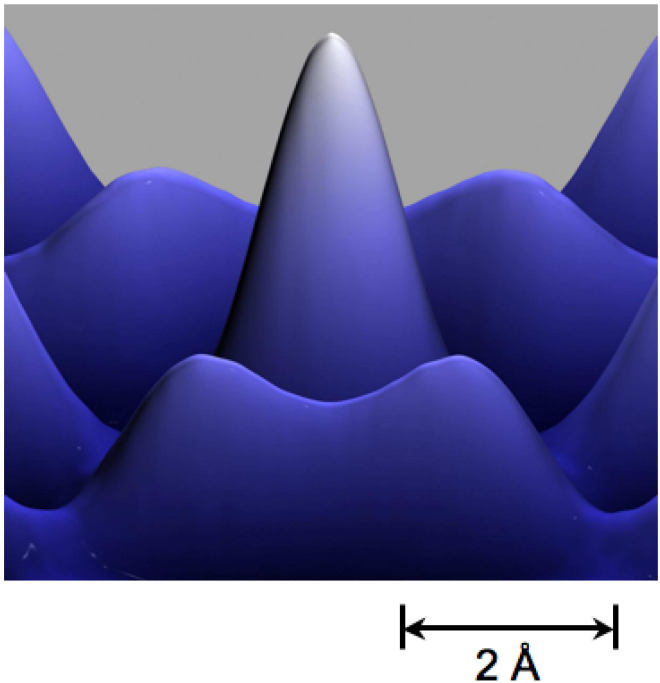,width=90mm,silent=}}
\caption{(Color online) Magnification of the area marked by the white, dashed 
square in Fig. 3. The width of this peak of the electron density is 0.92 {\AA}
(FWHM). }
\label{sentef4}
\end{figure}

As shown in Fig. \ref{sentef3}, 17 hard disks were placed on the contourline 
of a semi-ellipse with eccentricity $e=0.5$ and an average disk spacing of 6 
{\AA}. In this calculation the wavelength of the incoming wave of 3.2 {\AA} 
and a disk radius of again $r_0=0.63$ {\AA} was chosen. Fig. \ref{sentef3} 
shows the resulting contour plot of $|\psi({\bf r})|^2$ for the scattering 
state calculated from Eq. (\ref{psibe}). The complex structures in this 
interference pattern originate in part from the angular dependent p-wave 
scattering channels, which have no counterpart in classical geometrical 
optics. Fig. \ref{sentef4} shows a larger magnification of the area marked by 
the white dashed square in Fig. \ref{sentef3}. This area contains the most 
prominent constructive interference peak in this semi-elliptic focussing 
quantum mirror geometry. The peak has an anisotropic shape with an almost 
elliptic cross section; along the horizontal direction the FWHM of this peak 
is merely 0.92 {\AA}. This is less than 2 Bohr radii. The peak width is 
therefore smaller than the nominal size of the 1s orbital of hydrogen.

So far we have not made an attempt to uniquely determine the optimum disk 
arrangement, which leads to the sharpest interference structure. Alternative 
focussing geometries with different selected positions of the scattering 
centers may very well lead to even sharper interference peaks. Initial ideas of
``wave function engineering'' by a special-purpose design of quantum corral 
geometries have recently been formulated in the attempt to generate special 
predefined mirage phenomena \cite{Correa}. It is likely that similar 
strategies can be followed to identify arrangements of scattering centers with 
optimized focussing properties. If surface waves with wavelengths of just a 
few {\AA} are considered, such optimization strategies will necessarily have 
to include also non--$s$-wave scattering channels. 

Our calculations reveal in a proof-of-principle that special arrangements of 
individual atoms on surfaces allow to create electron states with diameters 
comparable to the size of a hydrogen atom. These states may be coupled to bulk
states and be used in devices such as highly focussed sources of tunneling 
electrons, as for example required for STM tips. The focussing of spin 
polarized surface states may furthermore allow to image magnetic structures on
the atomic or subatomic scale. The controlled design and device applications 
of electronic structures on the sub-{\AA} scale may therefore 
emerge as a real possibility. 

We thank F. Giessibl, D. Vollhardt, G. Sch\"on, {\O}. Fischer, M. Sekania, 
C. W. Schneider, T. Kopp, R. Claessen, and D. Pohl for thoughtful discussions. 
This work was supported by the BMBF (EKM-project 13N6918) and the Deutsche 
Forschungsgemeinschaft (SFB 484).


\begin{thebibliography}{99}
\bibitem{comment} Because we use the asymptotic form for the scattered waves, 
which is only very accurate on distances larger than 5 atomic radii, the 
amplitude for $\psi({\bf r})$ is overestimated in the very near vicinity of 
each scattering disk.
\end{thebibliography}
\end{document}